\documentclass[twocolumn,resetfootnote,longbib]{aastex701} 

\usepackage{graphicx}
\usepackage{float}
\usepackage{amsmath}
\usepackage{amssymb}
\usepackage{mathtools}
\usepackage{mathrsfs}
\usepackage{bm}
\usepackage{enumerate}
\usepackage{booktabs}
\usepackage[width=\columnwidth]{caption}

\newcommand{\Neost}{NEoST}
\newcommand{\Msun}{$M_\odot$}
\newcommand{\ntwolo}{N$^2$LO}
\newcommand{\nthreelo}{N$^3$LO}
\newcommand{\nsat}{
    \ifmmode n_0\else$n_0$\fi
}
\newcommand{\esat}{
    \ifmmode \varepsilon_0\else$\varepsilon_0$\fi
}
\newcommand{\chieft}{$\chi$EFT}

\graphicspath{{./}{figures/}}

\begin{document}

\title{Astrophysics Equation of State Inference with Bayesian Chiral Effective Field Theory Uncertainties}

\author[0000-0002-5250-0723]{Melissa Mendes}
\affil{Technische Universit\"at Darmstadt, Department of Physics, 64289 Darmstadt, Germany}
\affil{ExtreMe Matter Institute EMMI, GSI Helmholtzzentrum f\"ur Schwerionenforschung GmbH, 64291 Darmstadt, Germany}
\affil{Max-Planck-Institut f\"ur Kernphysik, Saupfercheckweg 1, 69117 Heidelberg, Germany}
\email{melissa.mendes@tu-darmstadt.de}

\author[0009-0005-7454-5956]{Hannah Göttling}
\affil{Technische Universit\"at Darmstadt, Department of Physics, 64289 Darmstadt, Germany}
\affil{ExtreMe Matter Institute EMMI, GSI Helmholtzzentrum f\"ur Schwerionenforschung GmbH, 64291 Darmstadt, Germany}
\email{hannah.goettling@tu-darmstadt.de}

\author[0009-0002-1785-9500]{Anna Hensel}
\affil{Technische Universit\"at Darmstadt, Department of Physics, 64289 Darmstadt, Germany}
\email{}

\author[0000-0002-9211-5555]{Isak Svensson}
\affil{Technische Universit\"at Darmstadt, Department of Physics, 64289 Darmstadt, Germany}
\affil{ExtreMe Matter Institute EMMI, GSI Helmholtzzentrum f\"ur Schwerionenforschung GmbH, 64291 Darmstadt, Germany}
\affil{Max-Planck-Institut f\"ur Kernphysik, Saupfercheckweg 1, 69117 Heidelberg, Germany}
\email{isak.svensson@tu-darmstadt.de}

\author[0000-0003-0640-1801]{Kai Hebeler}
\affil{Technische Universit\"at Darmstadt, Department of Physics, 64289 Darmstadt, Germany}
\affil{ExtreMe Matter Institute EMMI, GSI Helmholtzzentrum f\"ur Schwerionenforschung GmbH, 64291 Darmstadt, Germany}
\affil{Max-Planck-Institut f\"ur Kernphysik, Saupfercheckweg 1, 69117 Heidelberg, Germany}
\email{kai.hebeler@tu-darmstadt.de}

\author[0000-0001-8027-4076]{Achim~Schwenk}
\affil{Technische Universit\"at Darmstadt, Department of Physics, 64289 Darmstadt, Germany}
\affil{ExtreMe Matter Institute EMMI, GSI Helmholtzzentrum f\"ur Schwerionenforschung GmbH, 64291 Darmstadt, Germany}
\affil{Max-Planck-Institut f\"ur Kernphysik, Saupfercheckweg 1, 69117 Heidelberg, Germany}
\email{achim.schwenk@tu-darmstadt.de}

\author[0000-0002-9626-7257]{Nathan Rutherford}
\affil{Foundational Questions Institute (FQxI), 235 Ponce de Leon PI., Suite M, $\#217$, Decatur, GA 30030, USA}
\email{nathan.rutherford@fqxi.org}

\author[0000-0002-1009-2354]{Anna Watts}
\affil{Anton Pannekoek Institute for Astronomy, University of Amsterdam, Science Park 904, 1098 XH Amsterdam, The Netherlands}
\affil{Gravitation and Astroparticle Physics Amsterdam (GRAPPA), University of Amsterdam, 1098 XH Amsterdam, The Netherlands}
\email{A.L.Watts@uva.nl}

\begin{abstract}
We investigate Bayesian chiral effective field theory (\chieft) uncertainties, which assign a statistical interpretation to equation of state (EOS) distributions near nuclear saturation density,\nsat, as well as constraints from perturbative quantum chromodynamics (pQCD) to Bayesian EOS inference from LIGO/Virgo, NICER, and pulsar mass observations. The tails of the \chieft\ uncertainties allow for broader pressure ranges in our priors, but large parts of these are excluded by the astrophysical observations so that the EOS and the resulting mass-radius posteriors are still very consistent with our earlier work. Within our broad prior ranges, we observe a clear stiffening of the EOS at $n \gtrsim 3 n_0$. Moreover, the impact of the pQCD constraints on the posterior EOS and mass-radius range is negligible due to the astrophysics constraints. Exploiting the strong correlation between pure neutron matter and matter in beta equilibrium, we infer the symmetry energy slope parameter $L$ from astrophysics. For the $68\%$ credible interval, we obtain $L=42.6-52$\,MeV and $L=44.2-56.7$\,MeV using piecewise-polytrope and speed-of-sound high-density extensions, respectively. The $L$ posterior is mainly driven by the combination of GW170817 LIGO/Virgo and PSR J0740+6620, PSR J0437-4715, and PSR J0614-3329 NICER observations.
\end{abstract}


\section{Introduction}

Neutron stars are natural laboratories for studying the equation of state (EOS) of dense matter~\citep{Lattimer:2021emm,Chatziioannou:2024jsr}. Astrophysical observations of high-mass pulsars \citep{Demorest:2010bx,Antoniadis:2013pzd,Fonseca2016,Cromartie2020,Fonseca21} set lower limits on the maximum mass of neutron stars. Combined with
tidal deformabilities from LIGO/Virgo gravitational-wave (GW) measurements \citep{Abbott:gw170817,Abbott:gw190425} and NICER mass-radius information
\citep{Miller19,Riley19,Miller21, Riley21,Salmi2022,salmi_atmospheric_2023,Choudhury:2024xbk,dittmann_more_2024,Salmi:2024aum,salmi_J1231_2024,Vinciguerra:2023qxq,hoogkamer_cross-comparison_2025, Mauviard:2025dmd,Hoogkamer:2025,Miller:2025qfq,Qi2025,Kini:2026rjx}, this provides strong constraints on the EOS at supranuclear densities.

Going from the astrophysical data to the EOS typically involves the inclusion of nuclear physics constraints through the neutron star crust and the properties of neutron-rich matter around saturation density. This has benefited from advances using chiral effective field theory (\chieft) with powerful many-body calculations to reliably predict the EOS of neutron matter up to around $1.5 \nsat$ (with saturation density $\nsat = 0.16$\,fm$^{-3}$)~\citep{Hebeler:2013nza,Tews:2012fj,Lynn:2015jua,Drischler2019,Keller:2022crb,Tews:2024owl,Alp:2025wjn,Drischler:2026vdm} and matter in beta equilibrium with quantified uncertainties~\citep{Keller:2022crb,Gottling:2025ohe}. Beyond nuclear densities, this is combined with general high-density extensions that are agnostic of the particle composition and interactions, using, for example, piecewise polytropes~\citep{Read_2009,Hebeler:2013nza}, speed-of-sound parameterizations~\citep{Tews:2018kmu,Greif:2018njt}, piecewise speed-of-sound models~\citep{Brandes:2022nxa,Koehn:2025}, or Gaussian processes (GPs)~\citep{Essick:2020flb,Landry:2020vaw,Ng:2025,Gorda:2025aiu}, combined with Bayesian statistical frameworks to infer the EOS of neutron star matter.

In addition, GPs have recently been used to provide Bayesian estimates of the \chieft\ truncation uncertainties from order-by-order calculations around saturation density~\citep{Drischler:2020hwi_EOS,Drischler:2020yad_matter,Gottling:2025ohe}. Moreover, at larger densities, for $n \gtrsim 40-50 \nsat$, constraints on the EOS based on perturbative quantum chromodynamics (pQCD) calculations have been derived \citep{Gorda2021a,Gorda2021b,Komoltsev_2022}. Even though these densities are far beyond those reached in neutron stars, it remains an interesting question how pQCD constraints further constrain the EOS within the relevant density regime of neutron stars, as investigated in \cite{Komoltsev:2023,Koehn:2025}.

In this context, the goal of this work is to incorporate the new GP-based Bayesian \chieft\ uncertainties and constraints from pQCD in our neutron star Bayesian EOS inference framework, \Neost\ \citep{Raaijmakers:2025hbz}\footnote{See \url{https://github.com/xpsi-group/neost}. This work used \texttt{v2.3.0} of \Neost.}.
In particular, we will study the impact of the Bayesian \chieft\ uncertainties for matter in beta equilibrium from \cite{Gottling:2025ohe} at $0.5\nsat \leq n \leq 1.5\nsat$ and implement EOS extensions that are consistent with pQCD at extreme densities, following the procedure described in \cite{Komoltsev_2022}. Compared to our previous work \citep{Raaijmakers:2019dks,Raaijmakers:2021uju,Rutherford:2024srk,Mauviard:2025dmd}, the use of a \chieft\ distribution allows us to infer the slope parameter $L$ of the symmetry energy from astrophysical data and to test which observations are most constraining for $L$. The $L$ parameter has been studied from various EOS constructions and inferences (see, for example, \citet{Essick:2021,Hu:2021,Fearick:2023,Lattimer:2023}), and it also correlates with observables from nuclear experiments such as the neutron skin thickness.

This paper is organized as follows. In Section~\ref{sec:neost}, we briefly review the Bayesian EOS inference framework used. We discuss the introduction of the Bayesian \chieft\ uncertainties in Section~\ref{sec:gp} and the inclusion of pQCD-compatible extensions in Section~\ref{sec:pqcd}. In Section~\ref{sec:results}, we study the impact of these on the EOS and the resulting mass-radius posteriors based on gravitational-wave observations from LIGO/Virgo and pulse-profile modeling (PPM) from NICER, including the high-mass pulsar information. In particular, we also present results for the inferred symmetry energy slope parameter $L$. Finally, we summarize our main findings in Section~\ref{sec:conclusion}. Additional results are included in the Appendix for the incorporation of the new GP-based Bayesian \chieft\ uncertainties at next-to-next-to-leading order (\ntwolo). The data and plotting scripts needed to reproduce our results are available on Zenodo~\url{https://doi.org/10.5281/zenodo.21416452}.

\section{EOS inference framework}
\label{sec:neost}

We use the open-source EOS inference code \Neost~\citep{Raaijmakers:2025hbz}, which implements the Bayesian inference framework used in previous studies \citep{Greif:2018njt,Raaijmakers:2019qny,Raaijmakers:2019dks,Raaijmakers:2021uju,Rutherford:2024srk,Mauviard:2025dmd,Hoogkamer:2025}. Here, we only provide a brief overview of this framework.

\Neost~builds EOSs compatible with theoretical constraints at low densities. At densities up to $0.5\nsat$, we use the Baym-Pethick-Sutherland (BPS) crust EOS~\citep{Baym:1971pw}. The region $0.5\nsat < n \leq 1.5\nsat$ is described by a probability distribution derived from \chieft\ calculations of dense matter, covered in detail in Section~\ref{sec:gp}. The transition from BPS to \chieft\ is realized via linear interpolation from the last point in the crust to the first point of the \chieft\ EOS with an equal or higher pressure and density. Two different general high-density extensions are used beyond $1.5\nsat$: a piecewise-polytropic (PP)~\citep{Read_2009,Hebeler:2013nza} or a speed-of-sound (CS) model~\citep{Greif:2018njt}, both with the same parameter ranges as in \cite{Rutherford:2024srk} for the transition density $1.5 n_0$. The PP model has three segments of the form $P(n) = K \,(n/\nsat)^\Gamma$, with $\Gamma_1, \Gamma_2 \in [0,8]$ and $\Gamma_3 \in [0.5,8]$, where the first polytrope goes from $1.5 n_0$ to $n_1 \in [2,8.3] \, n_0$, the second segment goes from $n_1$ to $n_2 \in [2,8.3] \, n_0$, and the third goes from $n_2$ to the maximal central density. The CS model is given by (in units where the speed of light $c=1$)
\begin{equation}\label{eq:cs}
    c_{s}^2(x) =a_1 \mathrm{e}^{-\frac{1}{2}\left(x-a_2\right)^2 / a_3^2}+a_6+\frac{\frac{1}{3}-a_6}{1+\mathrm{e}^{-a_5\left(x-a_4\right)}} \,,
\end{equation}
where $c_s^2 = dP/d\varepsilon$, $x = \varepsilon/(m_{\mathrm{N}} n_0)$, and the nucleon mass $m_{\mathrm{N}}= 939.565$\,MeV. The parameters vary within the ranges of $a_1 \in[0.1,1.5]$, $a_2\in[1.5,12]$, $a_3/a_2\in[0.05,2]$, $a_4\in[1.5,37]$, and $a_5\in[0.1,1]$, and $a_6$ continuously matches to the \chieft\ EOS. As described in Section \ref{sec:pqcd}, all EOSs are checked for pQCD compatibility, and in case of failure, pQCD-compatible extensions are built. This ensemble of EOSs forms our prior, $p(\bm{\theta} | \mathbb{M}) \, p(\bm{\varepsilon} \,|\, \bm{\theta}, \mathbb{M})$, where $\mathbb{M}$ represents the EOS models, $\bm{\theta}$ represents its parameters, and $\bm{\varepsilon}$ represents its central energy densities. 

\begin{figure*}[t!]
\centering
\includegraphics[width=0.75\textwidth]{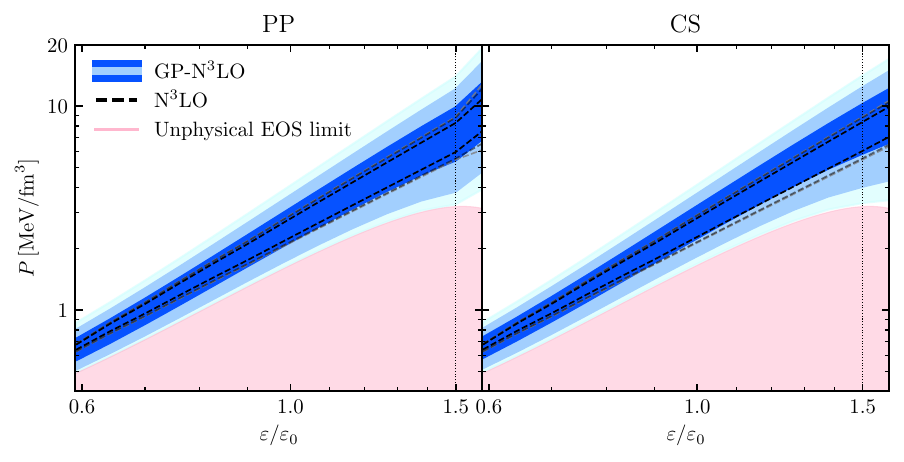}
\caption{Pressure-energy density prior distributions for the PP (left panel) and CS (right panel) extensions with Bayesian \chieft\ uncertainties up to $1.5\nsat$ (GP-\nthreelo, blue regions) and uniformly sampled in the previous \nthreelo\ range (\nthreelo, black dashed lines). The dark (light, lighter) blue regions and dashed black (gray, lighter gray) lines encompass the 68\% (95\% and 99.7\%) credible intervals for the GP-\nthreelo\ and \nthreelo\ cases, respectively. The pink region shows all EOSs excluded from the prior for their unphysical behavior above $1.5\esat$, indicated by the vertical thin dotted line.}
\label{fig:cEFT_PE_prior}
\end{figure*}

\Neost~uses Bayesian inference to find the most likely EOSs to describe given astrophysical data. Hence, with Bayes' theorem, the posterior distributions of the EOS can be written as
\begin{align}
p(\bm{\theta}, \bm{\varepsilon} \,|\, &\bm{d}, \mathbb{M})
\propto 
p(\bm{\theta} \,|\, \mathbb{M}) \,
p(\bm{\varepsilon} \,|\, \bm{\theta}, \mathbb{M}) \nonumber\\[1mm]
& \times \prod_{i} p(\Lambda_{1,i}, \Lambda_{2,i}, q_i \,|\, \mathcal{M}_c, \bm{d}_{\textnormal{GW}, i}) \nonumber\\
& \times \prod_{l} p(M_l, R_l \,|\, \bm{d}_{\textnormal{NICER+radio},l}) \,,
\label{eq:bayes}
\end{align}
where $\bm{d}$ represents the selected dataset, $\Lambda_{1,i}$ and $\Lambda_{2,i}$ are the tidal deformabilities associated with the GW data $\bm{d}_{\textnormal{GW}, i}$, and $\bm{d}_{\textnormal{NICER+radio},l}$ are the mass-radius ($M_l$--$R_l$) NICER data already combined with constraints from radio observations. As discussed in \cite{Raaijmakers:2021uju}, we fix the neutron star binaries' chirp mass, $\mathcal{M}_c$.
Our setup for the EOS parameters takes largely the same form as detailed in \cite{Rutherford:2024srk} and further used in \cite{Mauviard:2025dmd}, with two improvements explained in detail in Sections~\ref{sec:gp} and~\ref{sec:pqcd}. The astrophysics data entering the likelihood are as in \cite{Mauviard:2025dmd} and described in Section~\ref{sec:data}.

The EOS parameters are sampled using the nested sampling algorithm~\citep{Skilling04} implemented in \textsc{MultiNest}~\citep{Feroz09,Buchner14} using 5000 (100000) live points for calculating posteriors (priors). As of \texttt{v2.2.0}, \Neost\ utilizes the Message Passing Interface (MPI) multiprocessing feature of \textsc{MultiNest}, which substantially decreases the wall time of the sampling. MPI is further used in the postprocessing step of converting the EOS parameter samples to distributions of $M$, $R$, $P$, $\varepsilon$, $c_s^2$, and other EOS representations by solving the Tolman-Oppenheimer-Volkoff (TOV) equations.

\section{Bayesian \chieft\ uncertainties using\\ a Gaussian process}
\label{sec:gp}

Around nuclear saturation density, our EOS prior is based on order-by-order \chieft\ calculations of asymmetric nuclear matter by \cite{Keller:2022crb} using the N$^3$LO two- and three-nucleon (NN+3N) interaction with a cutoff of 450\,MeV \citep{Entem:2017gor,Drischler2019}. We consider a Bayesian approach to quantify truncation uncertainties of the EFT expansion, as introduced in \cite{Drischler:2020hwi_EOS,Drischler:2020yad_matter}, implemented by a two-dimensional GP in \cite{Gottling:2025ohe} that accounts for correlations along density and proton fraction. Here and in the following, we label our results obtained with this \chieft\ prior as ``GP-\nthreelo''.
We compare this EOS description to our previously used non-Bayesian model, known as the prescription of Epelbaum, Krebs, and Meißner (EKM)~\citep{Epelbaum:2014efa}, where the truncation uncertainty is estimated by inspecting expansion coefficients at lower orders and adopting the largest relative correction as the truncation error. We refer to results obtained with this method as ``\nthreelo''.

In our previous works \citep{Rutherford:2024srk,Mauviard:2025dmd}, we used the latter \nthreelo\ priors and assumed that the EOS was uniformly distributed in pressure-energy density ($P$--$\varepsilon$) space within the limits of the EKM prescription. Here, we improve on this by letting the EOS follow the normal distribution defined by the GP with the given mean and covariance functions from \cite{Gottling:2025ohe}. When drawing a sample from the GP representation of the pressure, we have to guarantee that it is monotonically increasing. This is not guaranteed for random samples drawn from the GP in the standard way. We ensure monotonicity by drawing $x$ from a normal distribution $\propto \exp(-x^2/2)$ to define a draw from the GP-\nthreelo\ distribution according to
\begin{equation}
    P_{\text{sample}}(n) = P_\mu(n) + x \cdot P_{\sigma} (n) \,,
\end{equation}
where $P_\mu(n)$ is the GP mean for matter in $\beta$ equilibrium at density $n$ and the standard deviation, $P_{\sigma}(n)$. The resulting GP-\nthreelo\ priors are shown in Figure~\ref{fig:cEFT_PE_prior}. At the 68\% credible interval, they are similar to the EKM prescription for the \nthreelo\ band but the tails of the GP uncertainties allow for broader pressure ranges.

As the truncation uncertainty encoded in the standard deviation is also a function of the density, it grows faster than the pressure itself at the tail of the distribution. Specifically for the lower end of the pressure probability at densities approaching $1.5\esat$ (where $\esat$ is the energy density at saturation), this behavior causes unphysical samples, with decreasing pressures. We exclude those samples before creating the high-density EOS extension. For GP-\nthreelo, this corresponds to assigning zero sampling probability to about $7\%$ of the EOSs at the lower pressure end of the populated interval of the distribution. This excludes the pink region in Figure~\ref{fig:cEFT_PE_prior} from the sampled GP-\nthreelo\ distribution. As a result, the 68\% region is slightly shifted to higher pressures for GP-\nthreelo\ versus \nthreelo.

\section{Enforcing pQCD compatibility}
\label{sec:pqcd}

\begin{figure}[t!]
\includegraphics[width=\columnwidth]{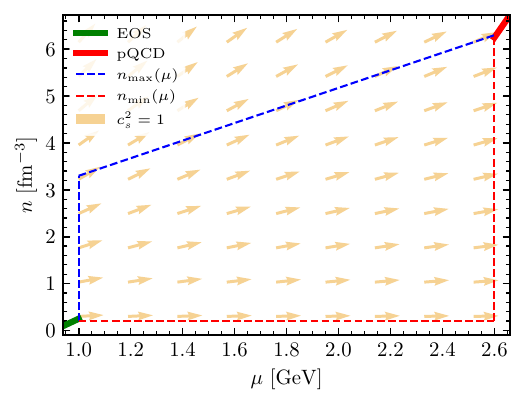}  
\caption{Visual representation of the ``minimum'' (red dashed line) and ``maximum'' (blue dashed line) pQCD-compatible extensions to an EOS (green) in the number density-chemical potential ($n$--$\mu$) plane. The pQCD limit is in red. Orange arrows indicate the $c_s^2=1$ lines.} 
\label{fig:pQCD_construction}
\end{figure}

We implement the compatibility with pQCD calculations at very high densities following the procedure described in \cite{Komoltsev_2022}\footnote{Code available at \url{https://github.com/OKomoltsev/QCD-likelihood-function}} and first employed in \cite{Gorda_2023}.
In \Neost, we check all EOSs with the PP or CS extensions for pQCD compatibility. For this procedure, a grid of EOS pressure-energy density points is checked on whether they could reach the calculated pQCD limits in a causal and thermodynamically consistent way. If one of these points fails the pQCD check before reaching the maximum stable neutron star central energy density, a pQCD-compatible extension is built from the last pQCD-compatible EOS pressure-energy density point on. This extension follows a ``minimum'' or ``maximum'' scenario, depending on whether the point failed the pQCD check because its pressure exceeded the upper or lower limit of the allowed pressure range. These extensions are illustrated in Figure~\ref{fig:pQCD_construction} in the $n$--$\mu$ plane. The pressure is connected to the $n$--$\mu$ plane by the pressure difference $\Delta P = \int_{\mu_{\mathrm{EOS}}}^{\mu_{\mathrm{pQCD}}}n(\mu) \mathrm{d}\mu$ between the last EOS point (green) and the first point of the pQCD limit (red). The red dashed line indicates the ``minimum'' extension scenario, where a luminal $(c_s^2=1)$ EOS is built from the last pQCD-compatible EOS point until the baryon chemical potential $\mu=2.6$\,GeV; then a phase transition in the $n$--$\mu$ plane is realized to the pQCD limit. The blue dashed line represents the ``maximum'' extension scenario, where the order of these processes is reversed and the phase transition is built first; then a luminal EOS connects to the pQCD limit.

The red pQCD limit shown in Figure~\ref{fig:pQCD_construction} was derived in~\cite{Gorda2021a,Gorda2021b}. Since the uncertainty of the pQCD calculations is dominated by truncation errors of the perturbative expansion, the results are dependent on the choice of renormalization scale $\Lambda$. To estimate this uncertainty, we work with the dimensionless renormalization scale $X= \frac{3\Lambda}{2\mu}$ and sample $X$ values according to a log-uniform distribution in the interval $[0.5,2]$, following the scale averaging discussed in~\cite{Gorda_2023}. The baryon chemical potential $\mu$ is fixed at $2.6$\,GeV~\citep{Gorda_2023,Fraga:2013}. At this value, the relative uncertainty turns out to be about $\pm 25 \%$ around the mean value. 

\begin{figure*}[t!]
\centering
\includegraphics[width=0.75\textwidth]{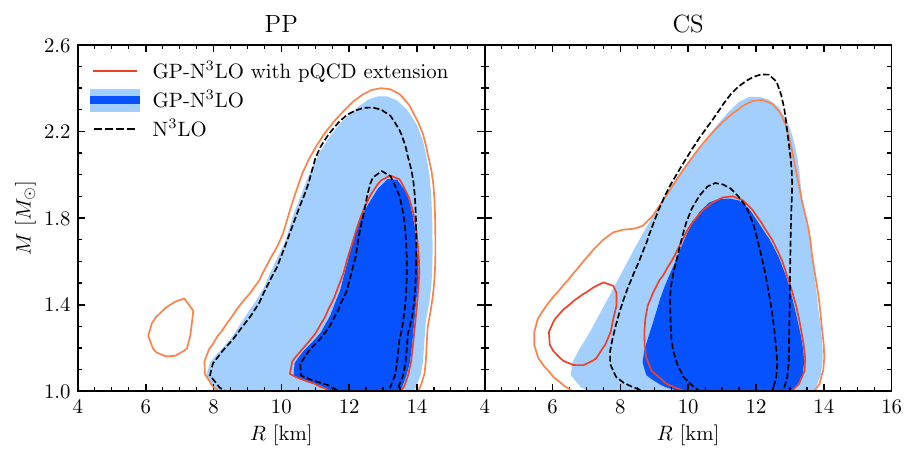}  
\caption{Mass-radius prior distributions for the PP (left panel) and CS (right panel) extensions with Bayesian \chieft\ uncertainties up to $1.5\nsat$ without pQCD extensions (GP-\nthreelo, blue regions), with pQCD extensions (GP-\nthreelo\ with pQCD extension, solid red  and orange lines), and uniformly sampled in the previous \nthreelo\ range (\nthreelo, black dashed lines). The dark (light) blue regions, the solid red (orange), and the inner (outer) black dashed lines encompass the 68\% (95\%) credible intervals. The regions at radii $R \lesssim 8\,$km in the orange/red contours are a result of softer EOSs being included in the prior by the pQCD extensions.}
\label{fig:MR_prior}
\end{figure*}

Including the pQCD extension during the EOS construction broadens the priors in the $P$--$\varepsilon$ and $M$--$R$ spaces. In particular, a significant number of soft EOSs are made compatible with the pQCD limit through ``minimum'' extensions, generating neutron stars with relatively low masses and radii (around $R \lesssim 8\,$km and $M \lesssim 1.6\,$\Msun), as can be seen in Figure~\ref{fig:MR_prior}. This is present when either GP-\nthreelo\ or \nthreelo\ EOSs are used (the latter is not shown). ``Maximum'' extensions are also present for stiffer EOSs, but they do not change the $P$--$\varepsilon$, $M$--$R$, or the maximum TOV mass-radius ($M_{\rm{TOV}}$--$R_{\rm{TOV}}$) prior distributions significantly, as the corresponding pQCD-incompatible points are already close to the maximum neutron star central densities. Thus, except for the excess at relatively low masses and radii, the impact of the pQCD extension is minor. As we will see, because these low masses and radii are excluded by astrophysical observations, the impact of the pQCD extension will become negligible for the posteriors. 

\section{Results of EOS inferences}
\label{sec:results}

In this section, we investigate the impact of the new GP-based \chieft\ prior and pQCD extensions on the EOS posteriors in the pressure-energy density, mass-radius, and speed-of-sound planes. We also infer the $L$ parameter, i.e., the slope of the symmetry energy, from astrophysical observations.

\subsection{Astrophysical data}
\label{sec:data}

Our input astrophysics data consist of four NICER pulsars and two gravitational-wave events from LIGO/Virgo. The pulsars, with their PPM configurations given in parentheses, are PSR J0740+6620 (\textsc{ST-U})~\citep{Salmi:2024aum}, PSR J0030+0451 (\textsc{ST+PDT})~\citep{Vinciguerra:2023qxq}\footnote{This analysis has recently been superseded by \cite{Kini:2026rjx}, which uses a larger NICER dataset together with XMM data and improved sampler settings. The results found in the newer study are, however, highly compatible with the \textsc{ST+PDT} results from \cite{Vinciguerra:2023qxq}.}, PSR J0437--4715 (\textsc{CST+PDT})~\citep{Choudhury:2024xbk}, and PSR J0614--3329 (\textsc{ST+PDT})~\citep{Mauviard:2025dmd}. For the first three, we employ the same configurations as in \cite{Mauviard:2025dmd}, and for PSR J0614--3329, we employ the headline result. Mass constraints from radio observations are included in the data in the cases of J0437, J0740, and J0614. In addition to the NICER data, we include the tidal deformabilities $\Lambda_1$ and $\Lambda_2$ and mass ratio $q$ of the two components of the gravitational-wave events GW170817~\citep{Abbott:gw170817} and GW190425~\citep{Abbott:gw190425}.

\subsection{Mass-radius posteriors}

\begin{figure*}[t!]
\centering
\includegraphics[width=0.8\textwidth]{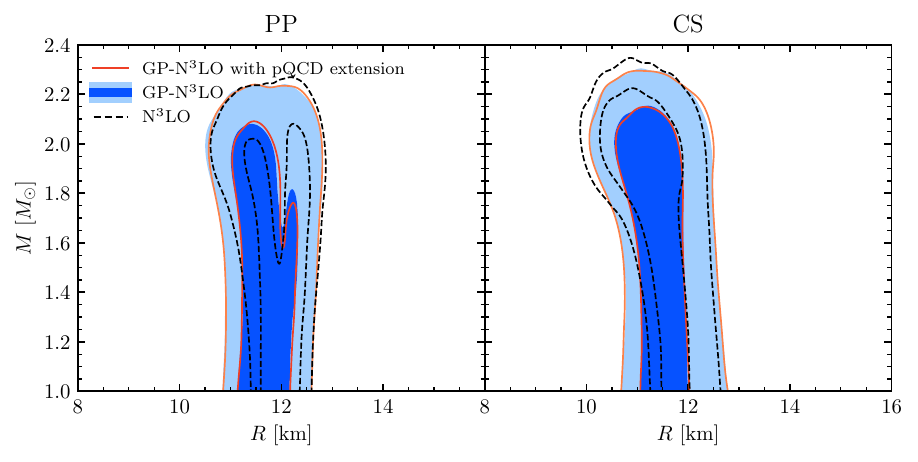}
\caption{Same as Figure~\ref{fig:MR_prior} but for the mass-radius posterior distributions.}
\label{fig:MR_posteriors}
\end{figure*}

\begin{figure*}[t!]
\centering
\includegraphics[width=0.8\textwidth]{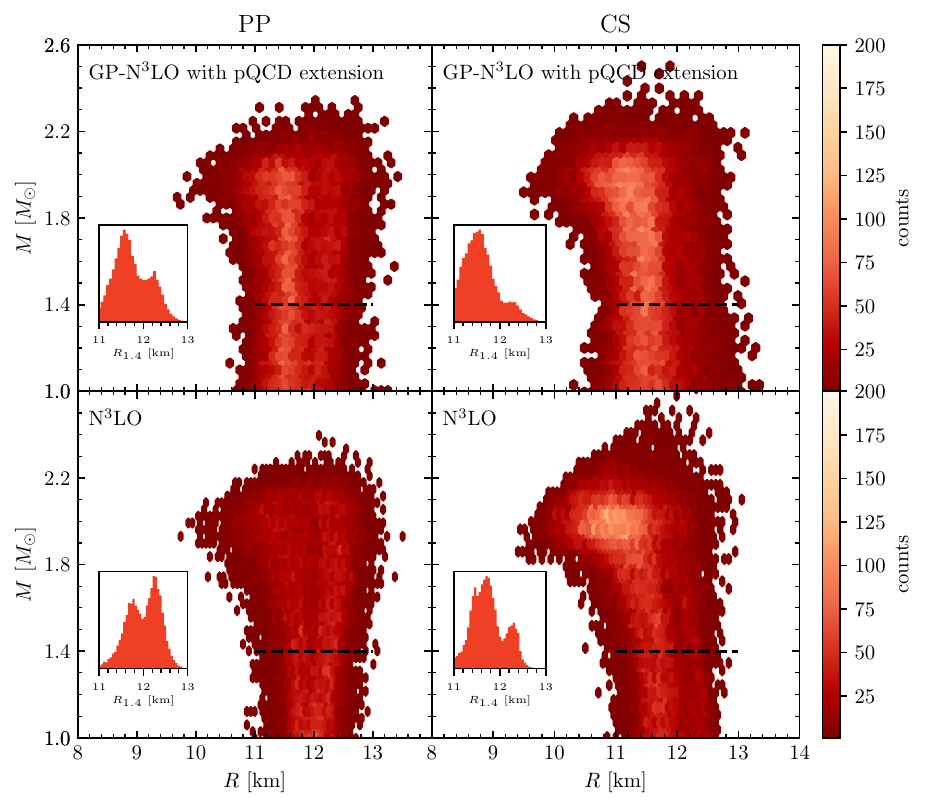}
\caption{Mass-radius posterior distributions showing the 2D histogram for the PP (left panel) and CS (right panel) extensions with Bayesian \chieft\ uncertainties up to $1.5\nsat$ with pQCD extensions (GP-\nthreelo\ with pQCD extension, upper panels) and uniformly sampled in the previous \nthreelo\ range (\nthreelo, lower panels). The dark shaded hexagons indicate a lower number of mass-radius samples, while the lighter shaded hexagons indicate a higher number of mass-radius samples. The insets show the radius distribution for a 1.4\,\Msun\ star, indicated by the black dashed lines.}
\label{fig:MR_posteriors_heatmap}
\end{figure*}

Figure~\ref{fig:MR_posteriors} reveals that the pQCD constraints have a negligible impact on the mass-radius posteriors for both PP and CS extensions. This shows that most of the EOSs included by pQCD extensions in the prior---primarily very soft EOSs---are excluded by the astrophysical data in the posterior. In particular, the presence of the high-mass pulsar J0740 disfavors these softer EOSs, explaining the disappearance of the EOSs at $R \lesssim 8\,$km and $M \lesssim 1.6\,$\Msun in the prior (see Figure~\ref{fig:MR_prior}).

\begin{figure*}[t!]
\centering
\includegraphics[width=0.8\textwidth]{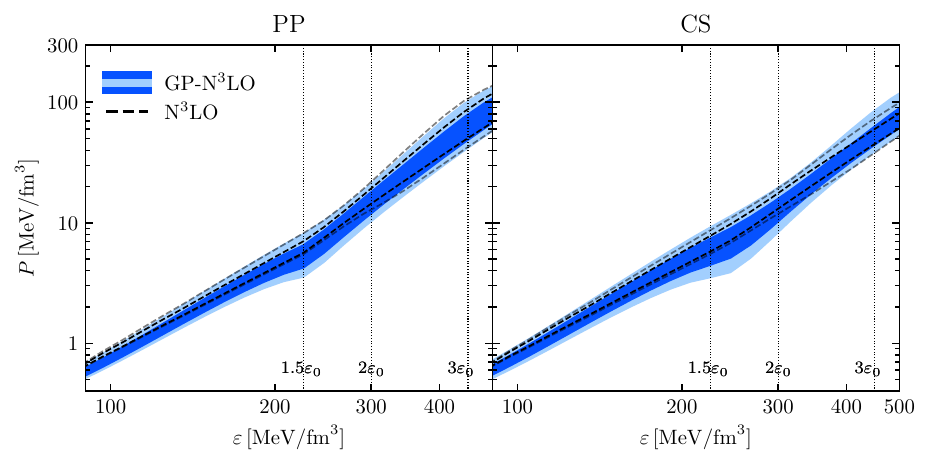}
\caption{Pressure-energy density posterior distributions for the PP (left panel) and CS (right panel) extensions with Bayesian \chieft\ uncertainties up to $1.5\nsat$ with pQCD extensions (GP-\nthreelo, blue regions) and uniformly sampled in the previous \nthreelo\ range (\nthreelo, black dashed lines). The dark (light) blue regions and the inner (outer) black dashed lines encompass the 68\% (95\%) credible intervals. The vertical thin lines mark the points of $1.5, 2$, and $3 \esat$ for easier reference.}
\label{fig:PE_J0614}
\end{figure*}

The new GP-based \chieft\ prior accommodates a wider range of EOSs in the low-density regime around saturation density than the uniformly distributed prior we have used in earlier works. It is therefore natural to expect wider posterior distributions, which we indeed observe. Figure~\ref{fig:MR_posteriors} shows that the mass-radius posteriors with GP-\nthreelo\ encompass a somewhat wider range of lower radii stars in the mass range $1.0-1.6$\,\Msun. At higher masses, the posteriors are largely unchanged: the PP GP-\nthreelo\ shifts weight slightly to low-radius high-mass stars, while the CS GP-\nthreelo\ posterior, slightly to high-radius high-mass stars.

Figure~\ref{fig:MR_posteriors_heatmap} shows the same results as Figure~\ref{fig:MR_posteriors}, but as two-dimensional histograms rather than credible intervals. This reveals some structure in the posteriors that is obscured in the plotting style of Figure~\ref{fig:MR_posteriors}. In particular, the bimodal-like structure around 1.4\,\Msun\ and heavier masses, previously seen in \cite{Rutherford:2024srk} and \cite{Mauviard:2025dmd}, is also present here. This structure is also visible to an extent in the left panel of Figure~\ref{fig:MR_posteriors}. This effect, where the posterior at $R \approx 12$\,km has a dip compared to both smaller and greater radii, may arise as a combination of stiffer priors and soft-favoring data (GW170817, J0614, and J0437) in combination with high-mass neutron stars with masses M$\geq2$\Msun (like J0740). 

\begin{figure*}[t!]
\centering
\includegraphics[width=\textwidth]{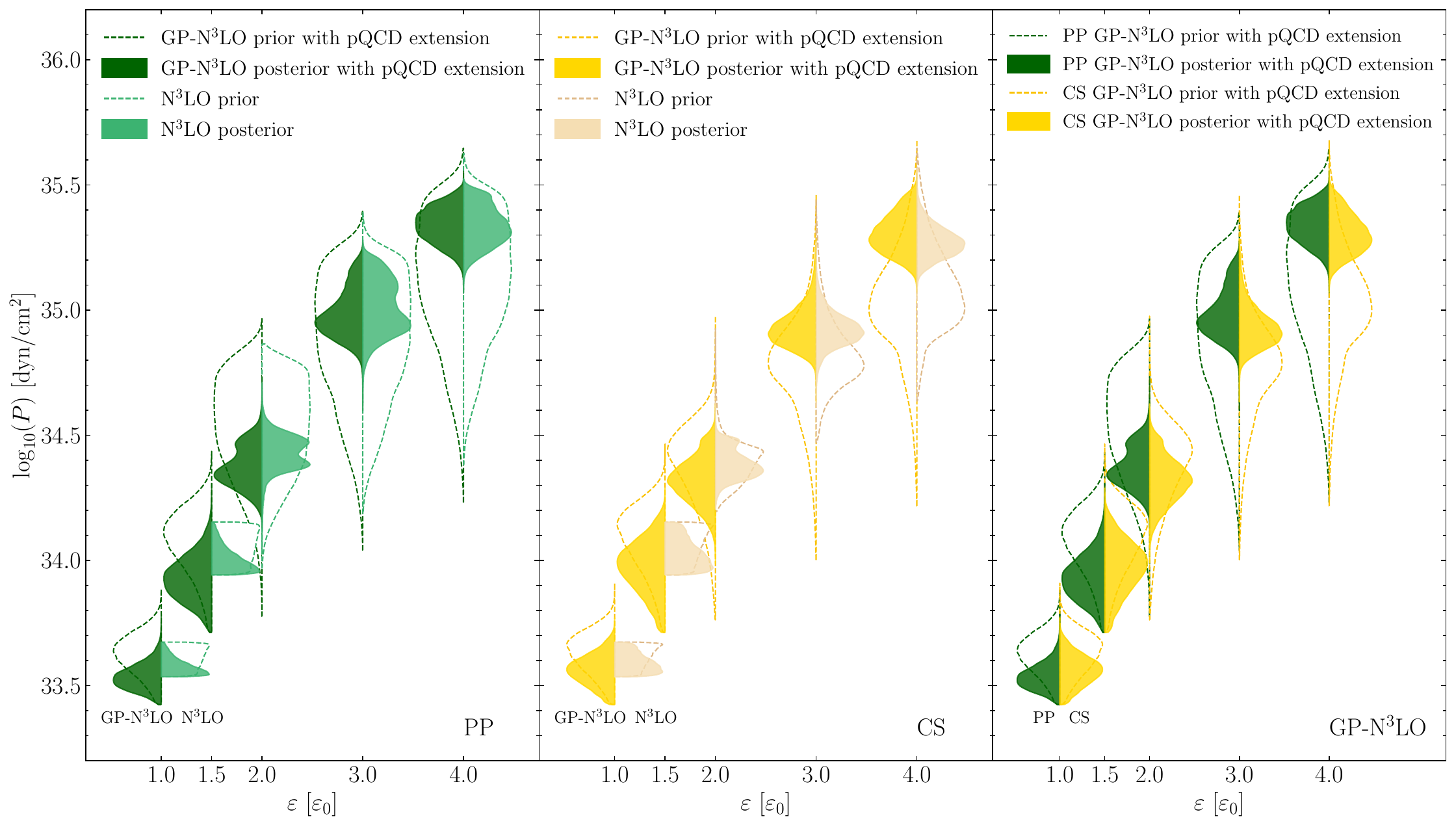}  
\caption{Prior (dashed lines) and posterior distributions (filled contours) for the pressure at $1, 1.5, 2, 3$, and $4 \esat$ with Bayesian \chieft\ uncertainties up to $1.5\nsat$ with pQCD extensions (GP-\nthreelo) versus uniformly sampled (\nthreelo) for the PP extension (left panel) and CS extension (middle panel) and a comparison between PP and CS extensions for GP-\nthreelo\ (right panel).}
\label{fig:violin_PE_J0614}
\end{figure*}

\begin{figure*}[t!]
\centering
\includegraphics[width=\textwidth]{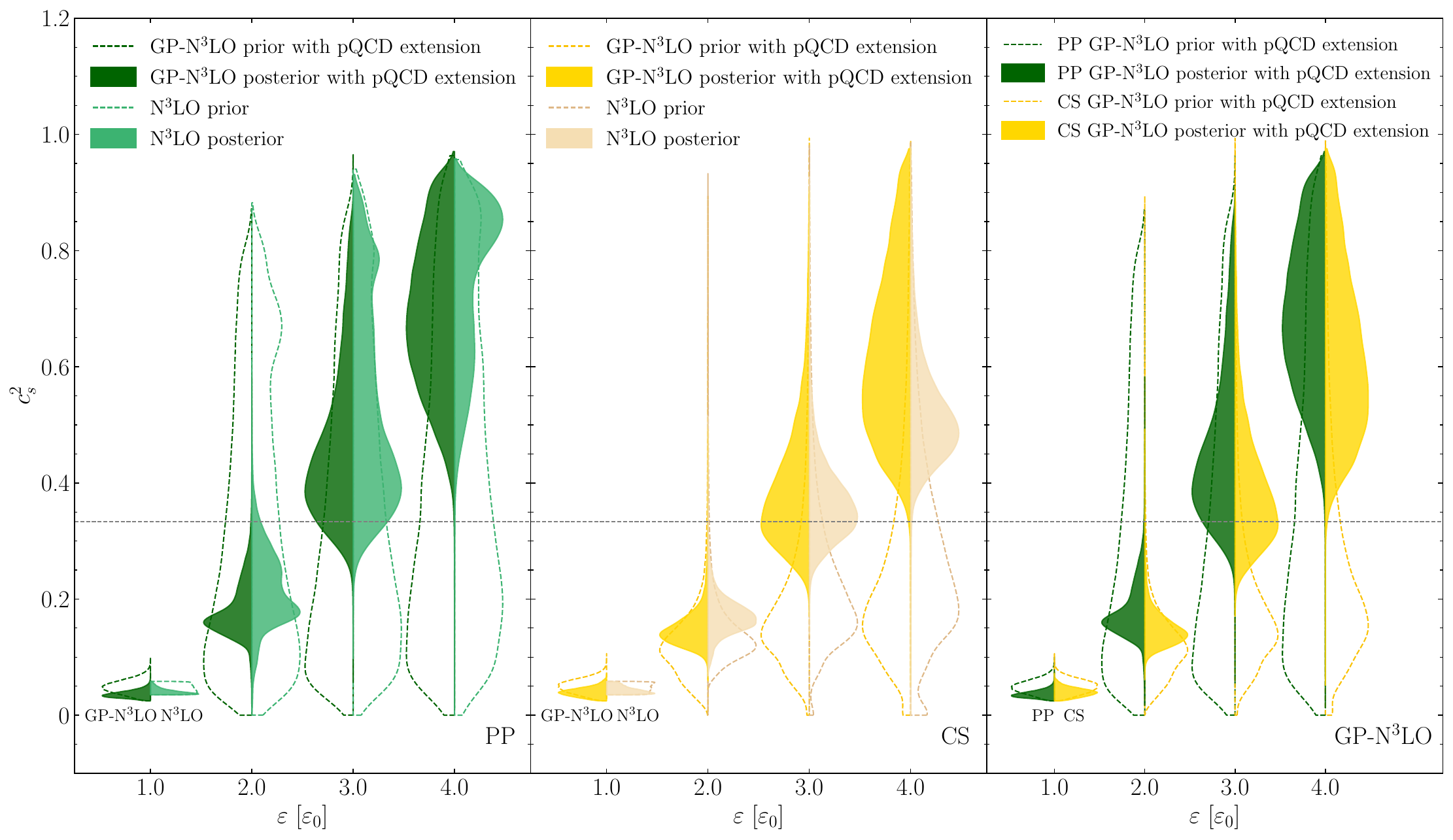}  
\caption{Same as Figure~\ref{fig:violin_PE_J0614} but for the speed of sound. The horizontal black dashed line shows the conformal limit $(c_s^2 = 1/3)$.}
\label{fig:violin_CS_J0614}
\end{figure*}

\subsection{Pressure-density posteriors}

In Figures~\ref{fig:PE_J0614} and \ref{fig:violin_PE_J0614}, the new $P$--$\varepsilon$ posteriors show a significant broadening to lower pressures below $2\esat$ as a direct result of using the GP-\nthreelo~\chieft. The upper limit of the posteriors is mostly unaffected, indicating that the astrophysical data already favor lower pressures around saturation density. This is clearly seen at $1 \esat$ in the leftmost panel of Figure~\ref{fig:violin_PE_J0614}, where the upper parts of the GP-\nthreelo\ and \nthreelo\ distributions are nearly identical, whereas lower pressures are cut off in the \nthreelo\ case. This effect, and the square shape of the prior for \nthreelo, are due to the uniform EOS sampling within the more restricted EKM limits. Nonetheless, beyond twice nuclear saturation density, the effect of the \chieft\ prior becomes very small. 

Most importantly, within our broad prior ranges, we observe a clear stiffening of the EOS at $\varepsilon \gtrsim 3 \esat$, from favoring softer parts of the prior to favoring stiffer parts in Figure~\ref{fig:violin_PE_J0614}. This stiffening is mostly driven by the presence of J0740, which requires neutron stars with a 2\,\Msun\ maximum mass. Similarly to the effects in the $M$--$R$ plane, discussed in the previous section, including pQCD-compatible extensions only slightly broadens the $P$--$\varepsilon$ priors, allowing for softer EOSs, and does not alter the posteriors visibly. We also note that the effect of the high-density extension (PP versus CS) is overall minor, with PP just showing a slight preference for higher pressures at $2-4 \esat$.

\subsection{Speed-of-sound posteriors}

Figure~\ref{fig:violin_CS_J0614} shows the speed of sound at $1, 1.5, 2, 3$ and $4 \esat$ corresponding to the pressures in Figure~\ref{fig:violin_PE_J0614}. Above saturation density, the GP-\chieft\ prior allows for more weight at lower speeds of sound around nuclear saturation density. Also, here, we observe a clear stiffening of the EOS between $2-3\esat$, with speeds of sound changing from mostly below the conformal limit, $c_s^2 = 1/3$, at $2\esat$ to mostly above at $3\esat$. While the posterior speed-of-sound range is similar for the PP and CS high-density extensions, the distributions are more different than for the pressure, especially at $4 \esat$. This is natural with the speed of sound being a derivative of the pressure. Also the differences of the GP-\nthreelo\ and \nthreelo\ cases are as expected from the pressure: while both have similar ranges, the GP-\nthreelo\ has more weight toward lower speeds of sound. 

\begin{deluxetable*}{l|ll|ll}[t!]
\tablewidth{5pt}
\tablecaption{Prior and Posterior Credible Intervals ($68\%$ and $95\%$) for the $L$ Parameter Based on the GP-\nthreelo\ EOS Extended until $1.5$\nsat\ with the PP and CS Extensions. In addition to the posterior ranges for the full LIGO/Virgo (GW170817 and GW190425) and NICER (J0740, J0437, J0614, and J0030), labeled ``Posterior'', we also give the ranges for the other astrophysics data scenarios considered in Figure~\ref{fig:L_multiple} and the prior (posterior) ranges based on the GP-\nthreelo\ EOS extended until $1.1$\nsat, labeled ``Prior (Posterior)$_{1.1n_0}$''.\label{tab:L}}
\tablehead{\colhead{Data Scenario} \vline& \multicolumn{2}{c|}{L [MeV] (PP)} & \multicolumn{2}{c}{L [MeV] (CS)}\\
\cline{2-5}
\colhead{} \vline& \colhead{$68\%$} & \colhead{$95\%$} \vline& \colhead{$68\%$} & \colhead{$95\%$}}
\startdata
Prior $_{1.1 n_0}$ & $47.7-67.7$ & $38.5-77.5$ & $47.4-68.1$ & $38.2-78.2$\\
Prior & $48.0-67.2$ & $40.5-75.9$ & $48.9-69.0$ & $40.5-76.5$\\
Posterior $_{1.1 n_0}$ & $37.2-56.9$ & $31.7-66.0$ & $41.6-58.0$& $35.1-66.4$\\
Posterior & $42.6-52.0$ & $40.0-58.6$ & $44.2-56.7$& $39.5-62.6$\\
J0740 & $51.0-66.3$& $45.6-74.7$ & $54.1-72.9$& $45.2-80.4$\\
J0740+J0437 & $47.1-60.4$& $43.2-68.9$ & $48.9-66.0$& $42.0-73.8$\\
J0740+GW170817 & $50.1-62.0$& $44.8-69.0$ & $52.5-68.3$& $43.4-74.0$\\
J0740+J0437+J0614+GW170817 & $43.1-53.3$ & $40.0-60.2$ & $43.5-56.3$& $39.5-63.5$\\
\enddata
\end{deluxetable*}

While both PP and CS extensions allow for first-order phase transitions (with $c_s^2 = 0$ regions), the astrophysical data disfavor these for the densities shown in Figure~\ref{fig:violin_CS_J0614}, as can be seen by the absence of a posterior excess at $c_s^2 = 0$. This is consistent with previous findings [see, for example, the previous study including both \chieft\ and pQCD constraints \citep{Gorda:2022lsk}].

\subsection{Inferring the $L$ parameter}
\label{sec:L}

With the sampled GP-\nthreelo\ distribution up to $1.5 n_0$, we can also infer the $L$ parameter of the EOS. The $L$ parameter characterizes the slope of the symmetry energy at saturation density $L = 3 n_0 \frac{\partial S(n)}{\partial n} \Bigr|_{n=n_0}$ and is thus related to the pressure of pure neutron matter. Given that neutron star matter and pure neutron matter are highly correlated---the correlation length of the two-dimensional GP in the proton-fraction ($x$) direction is $l_x = 0.31$~\citep{Gottling:2025ohe}---we can easily translate a draw from our normally sampled \chieft\ EOS for neutron star matter back to pure neutron matter by assuming they are fully correlated in the proton-fraction direction.

\begin{figure}[t!]
\centering
\includegraphics[width=\columnwidth]{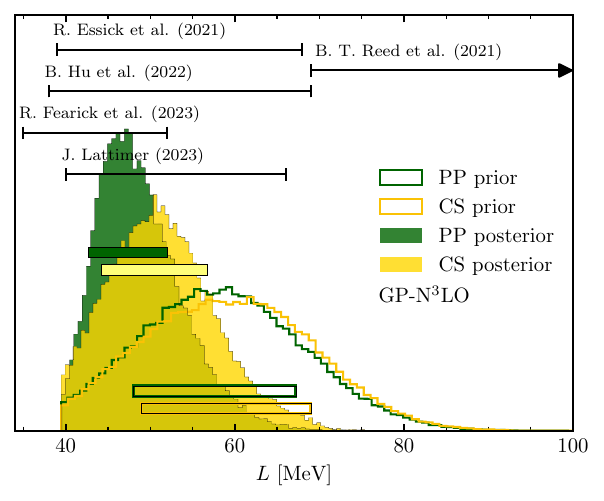}  
\caption{Prior (unfilled) and posterior (filled) distributions for the $L$ parameter based on the GP-\nthreelo\ EOS with the PP (green) and CS (yellow) high-density extensions. The posterior distributions include all astrophysics data considered in this work. We also give the $68\%$ credible interval as colored horizontal bars (unfilled prior and filled posterior), as well as comparisons to previous $L$ ranges (for details, see text in Section \ref{sec:L}; all intervals are 68\% except for \cite{Fearick:2023}, which is based on a set of NN+3N interactions). The upper extent of \cite{Reed:2021} is $143\,$MeV.}
\label{fig:L_inf}
\end{figure}

In Figure~\ref{fig:L_inf}, we show the prior and posterior distributions for the GP-\nthreelo\ EOS. The prior distributions reflect the \chieft\ uncertainties for the $L$ parameter based on the two-dimensional GP for the chiral NN+3N interaction used~\citep{Gottling:2025ohe}, and are thus identical for the PP and CS extensions except for the sampling statistics. The $L$ parameter ranges are consistent with the range for different chiral EFT interactions from \citet{Alp:2025wjn} given by $41.2-67.3$\,MeV. Note that the latter range does not include EFT truncation uncertainties, which make up the sampled GP-\nthreelo\ distribution based on the considered N$^3$LO NN+3N interaction \citep{Entem:2017gor,Drischler2019}. As shown by the limit of $L \gtrsim 39.5$\,MeV in Figure~\ref{fig:L_inf}, for lower $L$ parameters, our GP-\nthreelo\ EOS would lead to an unphysical EOS with a decreasing pressure at $1.5 n_0$ (see the discussion in Section~\ref{sec:gp}).

\begin{figure*}[t!]
\centering
\includegraphics[width=0.85\textwidth]{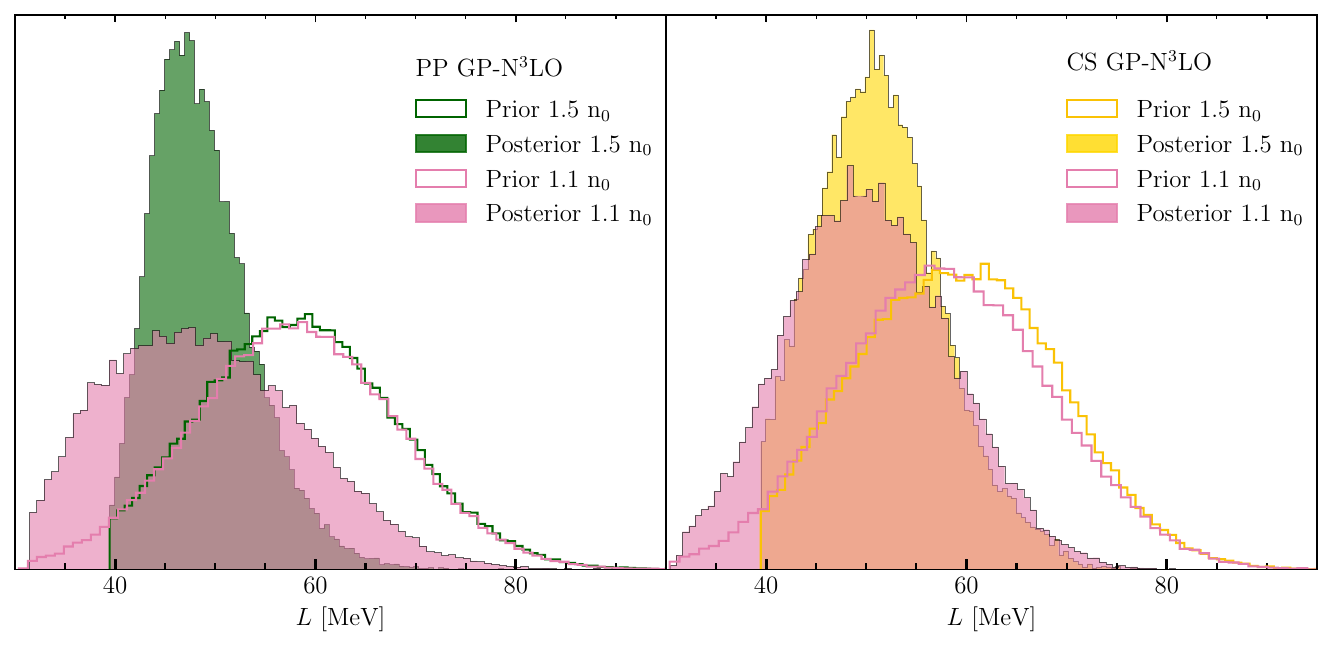}  
\caption{Prior (unfilled) and posterior (filled) distributions for the $L$ parameter for the GP-\nthreelo\ EOS extended up to 1.1 or 1.5\nsat\ with the PP (left) and CS (right) high-density extensions; the 1.5\nsat\ results are as in Figure~\ref{fig:L_inf}.}
\label{fig:L_1.1_1.5}
\end{figure*}

Our results for the posterior distributions for all astrophysics data are only weakly dependent on the high-density PP or CS extension, with the CS posterior extending to somewhat larger $L$. At the 68\% credible interval, we find the ranges for $L$ to be $42.6-52.0$\,MeV (PP) and $44.2-56.7$\,MeV (CS), preferring lower ranges than our priors. All 68\% and 95\% ranges for $L$ are also given in Table~\ref{tab:L}. As shown in Figure~\ref{fig:L_inf}, our posterior ranges for $L$ are also consistent with other astrophysics and/or $\chi$EFT-informed ranges from \cite{Essick:2021,Hu:2021,Fearick:2023,Lattimer:2023}, but we generally find stronger constraints from all astrophysics data considered in this work. Only the larger $L$ values obtained purely from PREX prefer larger $L$ values~\citep{Reed:2021} but also have very large uncertainties. We also note that our priors support high $L$ values, beyond the lower 68\% range of the PREX result \citep{Reed:2021}, but the astrophysics posteriors strongly disfavor $L > 70$\,MeV. In this context, we also mention a recent ab initio analysis of the PREX parity-violating asymmetry, which infers a smaller neutron skin \citep{Noel:2026byc}.

\begin{figure*}[t!]
\centering
\includegraphics[width=0.85\textwidth]{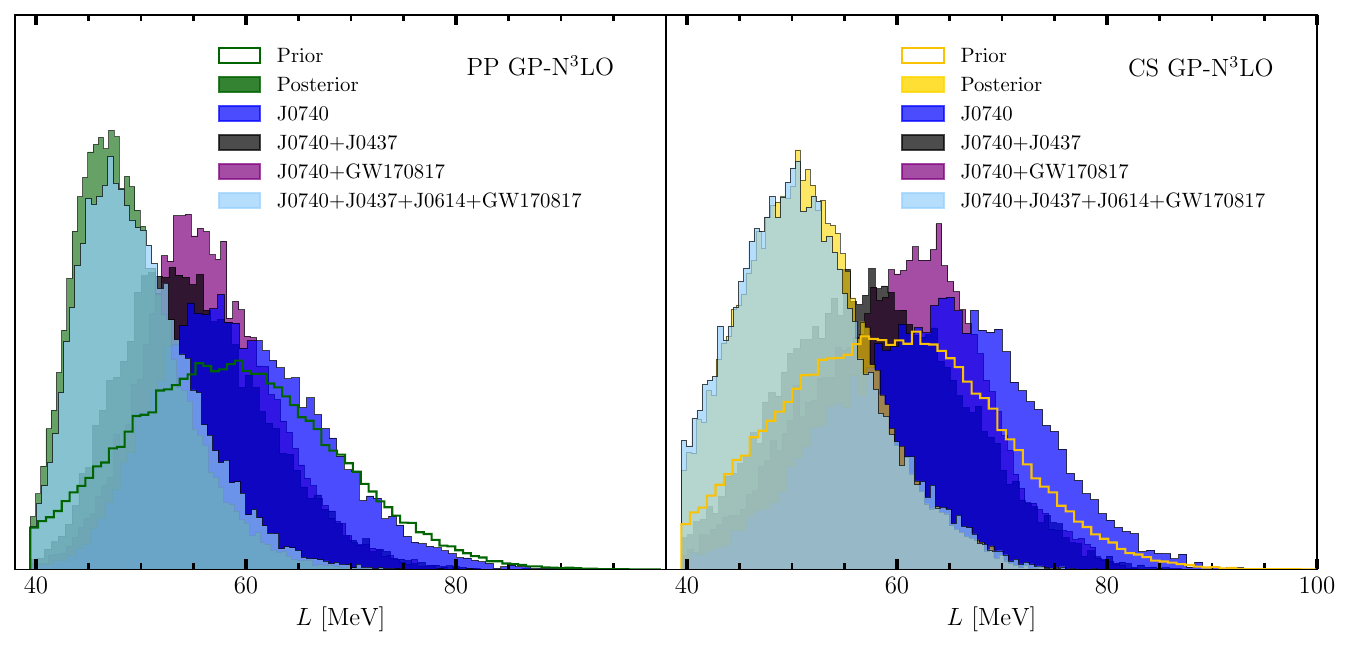}  
\caption{Distributions for the $L$ parameter for the GP-\nthreelo\ EOS with the PP (left) and CS (right) high-density extensions, as in Figure~\ref{fig:L_inf}, but based on different astrophysics data scenarios specified in the legend (for details see text in Section \ref{sec:L}).}
\label{fig:L_multiple}
\end{figure*}

The inference of the $L$ parameter can also be performed for a GP-\nthreelo\ EOS calculation extended until only $1.1 n_0$, as shown in Figure~\ref{fig:L_1.1_1.5}. In this case, the lower prior limit is $L \gtrsim 30.4$ MeV, as fewer EOSs reach unphysical values at this density, resulting in a less sharp low limit for $L$. Nonetheless, the $68\%$ credible intervals for $L_{1.1 n_0}$ and $L_{1.5 n_0}$ as well as the shape of the $L$ posterior are comparable; see also Table~\ref{tab:L}. Finally, we show in Figure~\ref{fig:L_multiple} the posterior distributions for $L$ obtained from different astrophysics data scenarios. This shows that the full astrophysics posterior is dominated by the combined data from ``J$0740$+J$0437$+J$0614$+GW$170817$'', while subsets of these observations are prior dominated and lead to a broader $L$ posterior extending to higher $L$ values. We thus expect the updated results for J0030 from \cite{Kini:2026rjx} to not significantly alter our posteriors. However, upcoming mass-radius contours from the high-mass pulsar PSR J1614-2230 could have an interesting impact. In particular, a smaller (larger) radius for this source than J0740 could pull $L$ to lower (larger) values, making this astrophysics information interesting for our analysis. We also studied the impact of including only the GW170817 or only the J0614 data, which also shifts the $L$ prior to smaller values, but both of these scenarios still present a significant overlap with the prior up to $L \approx 80$\,MeV.

\section{Summary and conclusions}
\label{sec:conclusion}

In this paper, we have studied the effects of two new developments for astrophysical EOS inferences: Bayesian \chieft\ uncertainty quantification and the incorporation of pQCD constraints. Using the GP from \citet{Gottling:2025ohe} to model the \chieft\ truncation uncertainties, we have built an EOS distribution around nuclear saturation density with a clear statistical interpretation. The GP allows for a significantly broader range of pressures up to about $1.5\nsat$ compared to our previous works. The resulting distributions include softer EOSs, while stiffer EOSs are disfavored by the astrophysical data. This led to a slight broadening of the $M$--$R$ posterior to lower radii for neutron stars with  $1.0-1.6$\,\Msun.

The enforcement of pQCD compatibility at extremely high densities following \citet{Komoltsev_2022} led to the inclusion of very soft EOSs in the prior. This was most notable in the $M$--$R$ prior, which exhibits a significant probability for low-mass stars with radii below 8\,km, whereas these stars were not included in the PP and CS ensembles before. However, the effects of the pQCD constraints on the posteriors were found to be negligible, as the vast majority of EOSs that are included through pQCD extensions are ruled out by the astrophysical data, mainly from the high-mass pulsar J0740.

Most interestingly for the EOS, we have found a clear stiffening of the pressure at $\varepsilon \gtrsim 3 \esat$, from favoring softer pressures at lower $\varepsilon$ to favoring stiffer EOSs above. This clear stiffening of the EOS between $2-3\esat$ was also seen in the speed of sound, with speeds of sound changing from mostly below the conformal limit, $c_s^2 = 1/3$, at $2\esat$ to mostly above at $3\esat$.

The strong correlation between pure neutron matter and matter in beta equilibrium has allowed us to also infer the symmetry energy slope parameter $L$ from astrophysics. For the $68\%$ credible interval, we obtained $L=42.6-52$\,MeV and $L=44.2-56.7$\,MeV---in the lower ranges of the prior---using the PP and CS extensions, respectively. Our findings are consistent with other theoretical and experimental inferences of the $L$ parameter, but our credible intervals are substantially narrower. Moreover, we have found that the $L$ posterior is mainly driven by the combination of GW170817 LIGO/Virgo and J0740, J0437, and J0614 NICER data.

LIGO/Virgo continues to gather data on GWs, and several other neutron star and GW observatories, such as Cosmic Explorer \citep{Reitze:2019}, Einstein Telescope \citep{Punturo:2010}, NewAthena \citep{Cruise:2025}, and eXTP \citep{Li:2025uaw}, are planned for the next decade. In addition, new radio facilities such as SKA~\citep{Basu:2025} promise important advances on the measurement of neutron star masses and moments of inertia. Upcoming data from these facilities will greatly increase our understanding of compact stars and the dense matter EOS. Our work shows how astrophysical data can directly inform the EOS behavior and nuclear matter parameters like $L$. These exciting prospects open up several avenues for future research, such as identifying whether compact star data EOS constraints are in agreement with laboratory data \citep{Somasundaram:2024ykk}. This kind of study also places high demands on theoretical EOS modeling, where all relevant sources of uncertainty need to be rigorously accounted for. Future work needs to also account for the uncertainties from the many-body method used to calculate the EOS at nuclear densities (see \cite{Svensson:2025jde} for initial steps) and the uncertainties from the description of the neutron star crust \citep{Gottling:2025ohe}. In addition, the agnostic high-density extensions need to be flexible enough to capture all possible EOS behaviors---degrees of freedom and interactions. To this end, GPs trained on various microphysics models \citep{Ng:2025} can help to validate that the prior does not bias the results. Finally, an interesting application of this work is to include the GP-\nthreelo\ in EOS-informed NICER data analysis \citep{Hoogkamer:2025}.

\section*{Acknowledgments}

We thank Tyler Gorda and Luis Hoff for useful discussions. This work was supported in part by the European Research Council (ERC) under the European Union's Horizon 2020 research and innovation programme (Grant Agreement No.~101020842) and by the LOEWE Top Professorship LOEWE/4a/519/05.00.002(0014)98 by the State of Hesse. We gratefully acknowledge the computing time provided on the high-performance computer Lichtenberg II at the TU Darmstadt. This is funded by the German Federal Ministry of Education and Research (BMBF) and the State of Hesse. We also gratefully acknowledge the Gauss Centre for Supercomputing e.V. (www.gauss-centre.eu) for providing computing time on the GCS Supercomputer JUWELS at Jülich Supercomputing Centre (JSC). N.R.~acknowledges generous support from the Foundational Questions Institute. A.L.W.~acknowledges support from NWO ENW-XL grant OCENW.XL21.XL21.038 {\it Probing the phase diagram of Quantum Chromodynamics}. 

\section*{Appendix}

\begin{figure*}[t!]
\centering
\includegraphics[width=0.75\textwidth]{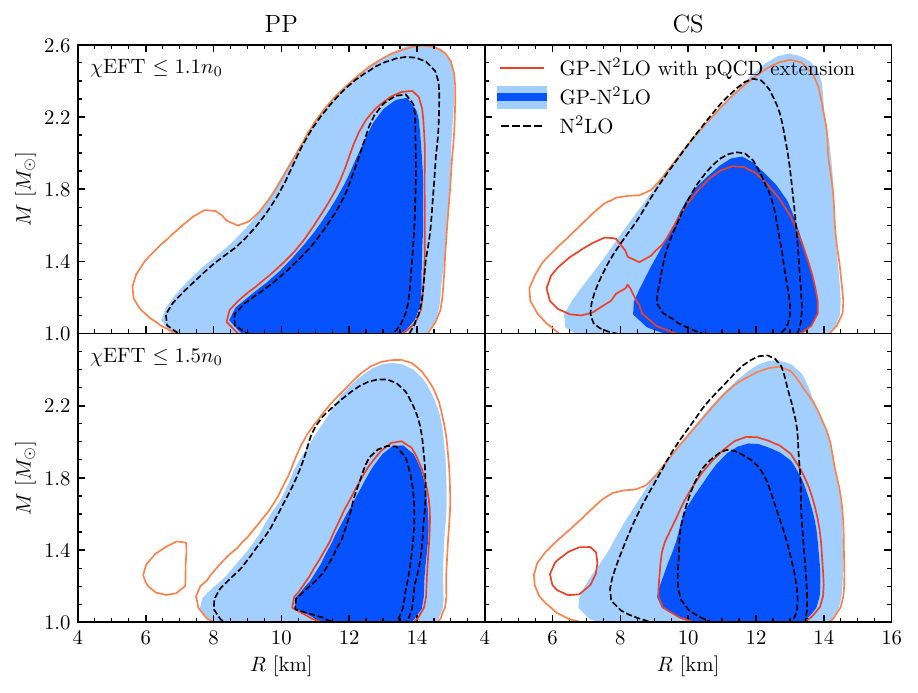}  
\caption{Same as Figure~\ref{fig:MR_prior} but with the \chieft\ uncertainties at \ntwolo\ up to $1.1\nsat$ (upper panels) and  $1.5\nsat$ (lower panels).}
\label{fig:MR_prior_n2lo}
\end{figure*}

\begin{figure*}[t!]
\centering
\includegraphics[width=0.75\textwidth]{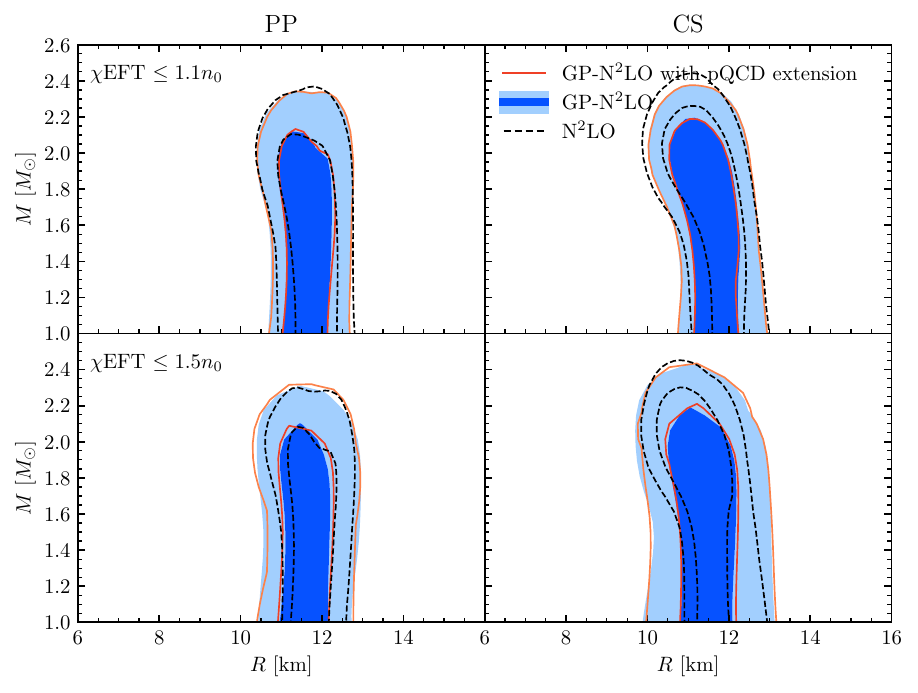}  
\caption{Same as Figure~\ref{fig:MR_posteriors} but with the \chieft\ uncertainties at \ntwolo\ up to $1.1\nsat$ (upper panels) and  $1.5\nsat$ (lower panels).}
\label{fig:MR_posterior_n2lo}
\end{figure*}

\begin{figure*}[t!]
\centering
\includegraphics[width=0.8\textwidth]{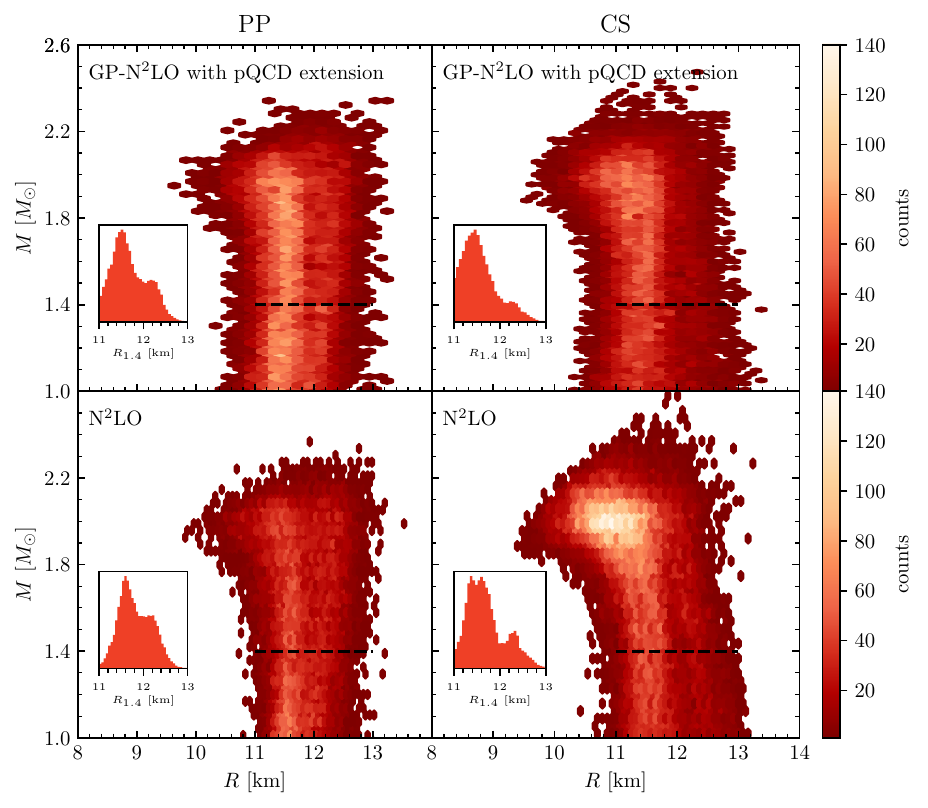}
\caption{Same as Figure~\ref{fig:MR_posteriors_heatmap} but with the \chieft\ uncertainties at \ntwolo.}
\label{fig:MR_posteriors_heatmap_n2lo}
\end{figure*}

\begin{figure*}[t!]
\centering
\includegraphics[width=0.8\textwidth]{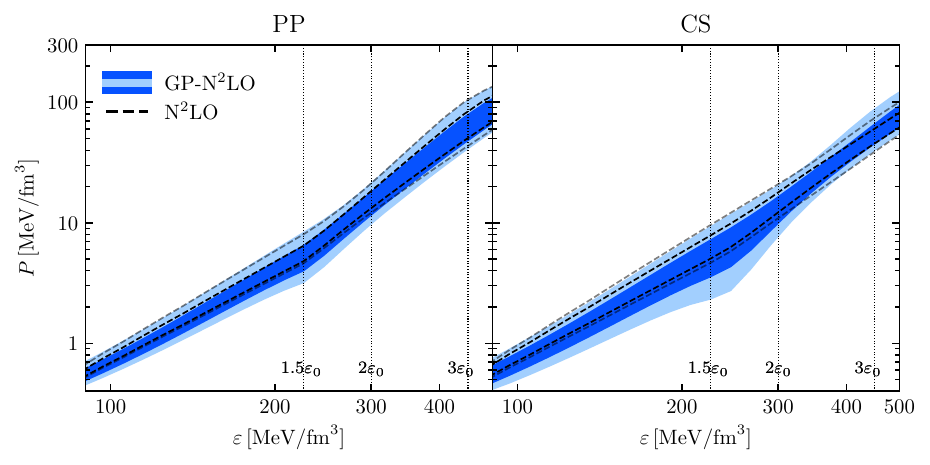}
\caption{Same as Figure~\ref{fig:PE_J0614} but with the \chieft\ uncertainties at \ntwolo.}
\label{fig:PE_J0614_n2lo}
\end{figure*}

\begin{figure*}[t!]
\centering
\includegraphics[width=\textwidth]{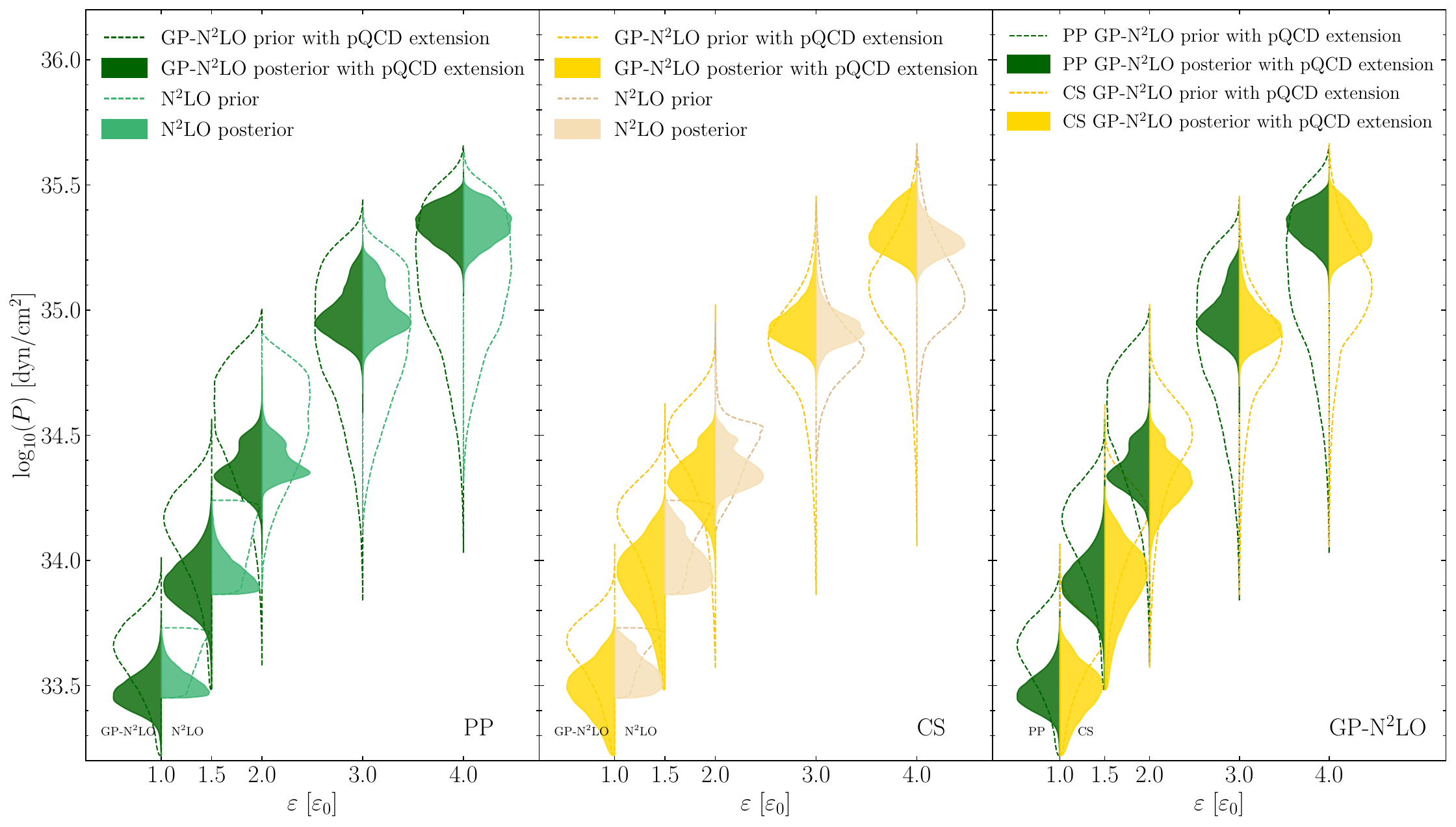}  
\caption{Same as Figure~\ref{fig:violin_PE_J0614} but with the \chieft\ uncertainties at \ntwolo.}
\label{fig:violin_PE_J0614_n2lo}
\end{figure*}

\begin{figure*}[t!]
\centering
\includegraphics[width=\textwidth]{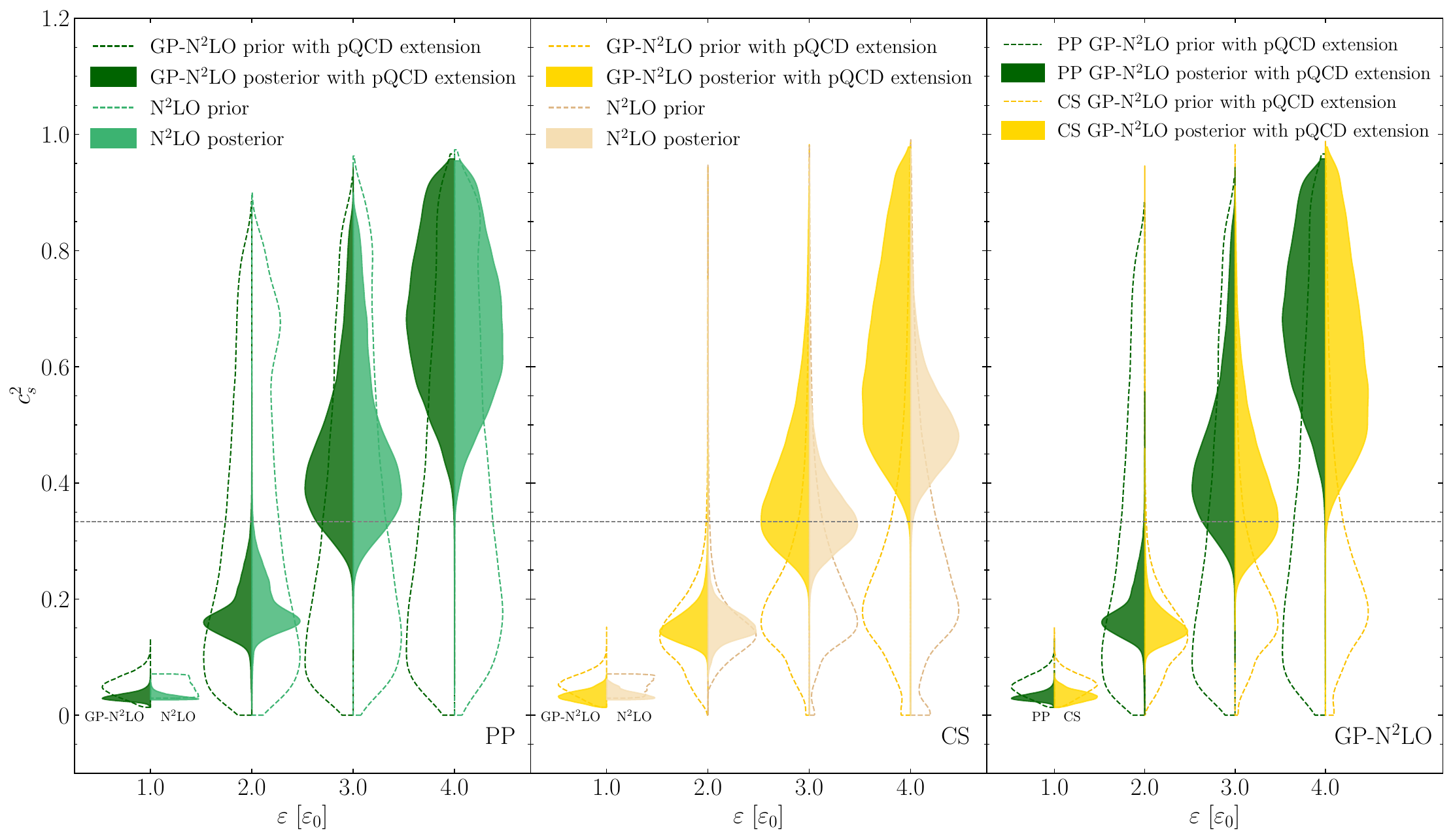}  
\caption{Same as Figure~\ref{fig:violin_CS_J0614} but with the \chieft\ uncertainties at \ntwolo.}
\label{fig:violin_CS_J0614_n2lo}
\end{figure*}

For completeness, we include here results when incorporating the new GP-based Bayesian \chieft\ uncertainties at \ntwolo\ in comparison to the previous \ntwolo\ results \citep{Rutherford:2024srk} in Figures~\ref{fig:MR_prior_n2lo},~\ref{fig:MR_posterior_n2lo}, and~\ref{fig:MR_posteriors_heatmap_n2lo} for the mass-radius distributions, in Figures~\ref{fig:PE_J0614_n2lo} and \ref{fig:violin_PE_J0614_n2lo} for the pressure-energy density distributions, and in Figure~\ref{fig:violin_CS_J0614_n2lo} for the speed-of-sound distributions.

\bibliography{lit}
\bibliographystyle{aasjournalv7}

\end{document}